\documentclass[aip,pof,longbibliography,twocolumn,reprint,groupedaddress]{revtex4-1}
\usepackage{graphicx}
\usepackage{float}
\usepackage{siunitx}
\usepackage{cancel}

\usepackage{bm,amssymb,amsmath,amsthm,mathtools,xcolor,array}

\definecolor{darkgreen}{rgb}{0,0.6,0.0}

\newcommand{\VEC}[1] {\mathbf{#1}}
\newcommand{\HAT}[1] {\hat{\mathbf{#1}}}

\newcommand{\derpar}[2]{\frac{\partial #1}{\partial #2}}
\newcommand{\Dr}{D_\text{r}}
\newcommand{\Pe}{{\rm Pe}}

\begin{document}

\title{Self-diffusive dynamics of active Brownian particles at moderate densities}

\author{Rodrigo Soto}
\email{rsoto@uchile.cl}
\affiliation{Departamento de F\'{\i}sica, Facultad de Ciencias F\'{\i}sicas y Matem\'aticas, Universidad de Chile,  Santiago, Chile}

\date{\today}

\begin{abstract}
The Active Brownian Particle  (ABP) model has become a prototype of self-propelled particles. ABPs move persistently at a constant speed $V$ along a direction that changes slowly by rotational diffusion, characterized by a coefficient $\Dr$. Persistent motion plus random reorientations generate a random walk at long times with a diffusion coefficient that, for isolated ABPs in two dimensions, is given by $D_0=V^2/(2\Dr)$. Here we study the density effects on the self-diffusive dynamics using a recently proposed kinetic theory for ABPs, in which persistent collisions are described as producing a net displacement on the particles. On intermediate timescales, where many collisions have taken place but  the director of the tracer particle has not yet changed, it is possible to solve the Lorentz kinetic equation for a tracer particle. It turns out that, as a result of collisions, the tracer follows an effective stochastic dynamics, characterized by an effective reduced streaming velocity $V_\text{eff}$ and anisotropic diffusion, with coefficients explicitly depending on the density. Based on this result, an effective theoretical and numerical approach is proposed in which the particles in a bath follow stochastic dynamics with mean-field interactions based on the local density. Finally, on time scales larger than $\Dr^{-1}$, studying the van Hove function at small wavevectors, it is shown that the tracer particle presents an effective diffusive motion with a coefficient $D=V_\text{eff}^2/(2\Dr)$. The dependence of $V_\text{eff}$ on the density indicates that the kinetic theory is limited to area fractions smaller than 0.42, and beyond this limit unphysical results appear.\end{abstract}

\maketitle


\section{Introduction}

Self-propelled particles such as swimming bacteria, mesenchymal cells, active colloids, or microrobots move persistently along directions that change randomly on long time scales~\cite{bechinger2016active}. Bacteria and other flagellated microorganisms change their director in tumbling events, where a new director is randomly chosen to start a new run phase~\cite{berg1993random,Berg2001E.ColiMotion}. In mesenchymal cells, active direction reversal occurs when the cell comes into contact with another cell in a process called contact inhibition of locomotion~\cite{stramer2017mechanisms}. Active colloids and microrobots normally change their direction continuously only by rotational diffusion~\cite{paxton2004catalytic,bricard2013emergence}. In fact, rotational diffusion is also present in swimming bacteria and  mesenchymal cells, which is superimposed on the other active reorientation processes. In general, the intensity of  rotational diffusion need not  be thermal, as other processes may be in action. In any case, the combination of persistent motion and random reorientations generates an effective random walk for isolated swimmers. Dimensional analysis shows that the diffusion coefficient should scale as $D_0\sim V^2/\tau$, where $V$ is the self-propulsion speed and $\tau$ is the average persistence time. The quantitative calculation of the prefactor has been accurately obtained for run-and-tumble bacteria~\cite{berg1993random,Berg2001E.ColiMotion}, bacteria with long internal chemotactic memory~\cite{villa2020run}, active colloids~\cite{howse2007self}, and several extensions of these models~\cite{lindner2008diffusion,romanczuk2012active,kumar2021effect,szamel2024extremely}.
 
Since  rotational diffusion is a general mechanism present in self-propelled particles, it becomes relevant to isolate this mechanism to study its effects. The Active Brownian Particle (ABP) model has emerged as a prototype of self-propelled particles where the director changes only by rotational diffusion and the particles interact only through excluded volume~\cite{romanczuk2012active,cates2013active,solon2015active}. In its simplest form, the position $\VEC{r}_i$ and the director $\HAT{n}_i$ of the $i$-th particle evolve as
\begin{align}\label{eq.ABP}
\dot{\VEC{r}}_i &= V \HAT{n}_i+\VEC{F}_i/\gamma, \\
\dot{\HAT{n}}_i &= \sqrt{2\Dr}\,\bm{\xi}_i(t)\times\HAT{n}_i,
\end{align}
where $\bm{\xi}_i$ are vectorial white noises of correlation $\langle \xi_{\mu,i}(t)\xi_{\nu,j}(t')\rangle=\delta_{\mu,\nu}\delta_{i,j}\delta(t-t')$, with $\mu,\nu=x,y,z$, the Cartesian coordinates. Here, $\VEC F_i=-\derpar{U}{\VEC r_i}$ is the hard-core interparticle force acting on $i$, where $U$ is the potential energy, $\gamma$ is the mobility coefficient, which is assumed to be constant, and $\Dr$ is the rotational diffusion coefficient. The dynamics of ABPs is non-inertial, reflecting the motion of microscopic particles moving at low Reynolds number.

In the following, although much of the discussion applies to arbitrary spatial dimensions, we will concentrate on the two-dimensional case, which is experimentally relevant since active colloids are usually non-buoyant and sediment to the bottom of the observation box. In this case, the director is simply characterized by an angle for each particle $\HAT n_i=(\cos\phi_i,\sin\phi_i)$, which evolves as
\begin{align}\label{eq.ABP2D}
\dot{\phi}_i &= \sqrt{2\Dr}\,\xi_i(t),
\end{align}
with $\xi_i$  white noises of correlation $\langle \xi_{i}(t)\xi_{j}(t')\rangle=\delta_{i,j}\delta(t-t')$.
In the limit of hard-core interactions, with diameter $\sigma$, a collection of $N$ ABPs moving in a two-dimensional box of area $A$ is characterized by two dimensionless parameters: the area fraction $\eta=\pi\sigma^2\rho_0/4$, where $\rho_0=N/A$ is the number density, and the P\'eclet number  $\Pe=V/(\sigma\Dr)$, which quantifies the persistence of the particles. More specifically, $\Pe$ compares the time it takes for the particle to change direction, $\Dr^{-1}$, with the time it takes for a particle to travel its size, $\sigma/V$.

\begin{figure*}[htb]
\includegraphics[width=2\columnwidth]{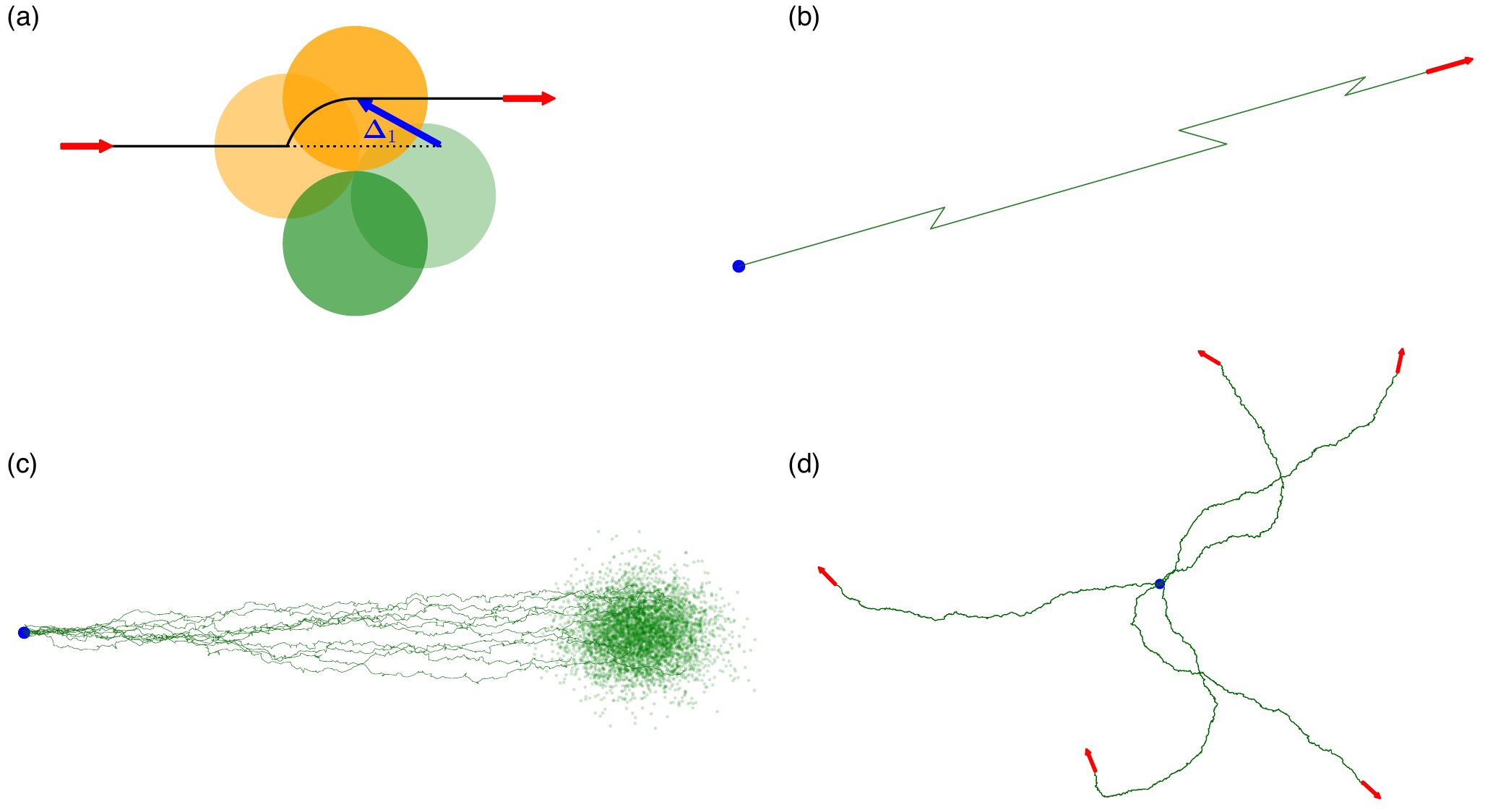}
\caption{Schematic representation of the particle dynamics at different timescales. The red arrow represents the director and the blue circle the initial position. 
(a) Elementary collision process in which a particle moving to the right (orange) collides with another particle (green). The solid black line shows the actual trajectory and the dotted black line shows the trajectory in absence of the collision. The blue arrow shows the effective displacement $\VEC \Delta_1$ produced by the collision. 
(b) Effective trajectory after a few collisions for $t_\text{col},t_\text{free}\lesssim t \ll t_\text{rot}$. 
(c) Effective Brownian dynamics for $t_\text{col},t_\text{free}\ll t \ll t_\text{rot}$, implying that the director is fixed pointing to the right. Example trajectories, obtained by numerically solving Eq.~\eqref{eq.effectiveLangevin} with $\Dr=0$,  are shown in dark green, and the light green dots indicate the final positions, where it can be seen that  $D_\parallel>D_\perp$. 
(d) Effective Brownian dynamics valid for $t_\text{rot}\lesssim t$, allowing the director to change. The sample trajectories are obtained by numerically solving Eq.~\eqref{eq.effectiveLangevin} with a finite value for $\Dr$.}
\label{fig.figure}
\end{figure*}

A quite striking feature of ABPs is that, for sufficiently large densities and persistence, they exhibit a phase separation forming large clusters, even though their interaction is purely repulsive~\cite{fily2012athermal,redner2013structure,levis2014clustering,fily2014freezing,digregorio2018full}. The origin of this motility induced phase separation (MIPS) is that in a dense medium,  particles tend to move at a reduced velocity due to collisions. The velocity reduction depends on density and  can trigger a density instability~\cite{tailleur2008statistical,cates2015motility}. Recently, a kinetic equation has been proposed to describe the dynamics of a collection of ABPs at high P\'eclet number, correctly predicting the area fraction to generate MIPS based  on  purely microscopic arguments~\cite{PRL}.  The idea behind this kinetic theory is that collisions take a finite time due to persistence. If $\Pe\gg1$, the directors do not change during the encounter, but the collision has a net effect of displacing the particles (see Fig.~\ref{fig.figure}a). These displacements are obtained as the difference between the actual positions at the end of the collision and the positions they would have had in the absence of the event.

Despite the importance of the model, the diffusion process for dense ABP systems has received little attention. Ref.~\onlinecite{wang2019anomalous} presents numerical results on the density dependence of the self-diffusion coefficient. As an effect of crowding, it shows a decreasing tendency with increasing density. Using a mode-coupling theory, the self-diffusion coefficient is computed theoretically and it also shows that it decreases with density~\cite{reichert2021tracer}. In both cases, however, no expression for the density dependence is given.

In this article, we use the kinetic theory discussed above to compute the diffusive dynamics in moderately dense systems. Special attention is given to finding explicit expressions for the density dependence of the various transport coefficients. The article is organized as follows. 
Section~\ref{sec.kt} revisits the kinetic equation derived in Ref.~\cite{PRL} and discusses the possible approximations for the pair correlation function at contact. The kinetic equation for a tagged particle moving in a system of ABPs is derived in Sect.~\ref{sec.kttagged} and the different time regimes are introduced. The analysis of the self-diffusion process of the tagged particle is performed in Sects.~\ref{sec.intermediate} and \ref{sec.long} for the intermediate and long time regimes, respectively. In both cases, the effective diffusion coefficients and the effective streaming velocity are obtained. Finally, the conclusions are presented in Sect.~\ref{sec.conclusions}.

\section{Kinetic description for a gas of ABPs} \label{sec.kt}

In the kinetic language, a collection of ABPs is described by the distribution function $f(\VEC r_1,\HAT n_1,t)$, which gives the number of particles in the vicinity of  $\VEC r_1$ and director $\HAT n_1$, at time $t$. The distribution function is normalized so that 
\begin{align}
\rho(\VEC r_1,t) = \int f(\VEC r_1,\HAT n_1,t) d\HAT n_1
\end{align}
is the number density. In Ref.~\onlinecite{PRL}, it is shown that the evolution of $f$ is given by a Boltzmann-like equation
\begin{equation}\label{eq:kin}
    \frac{\partial f}{\partial t} + V\mathbf{\HAT{n}}_1\cdot\derpar{}{\VEC r_1} f = D_\text{r}\nabla^2_{\HAT{n}_1} f + J[f],
\end{equation}
where the left-hand side (l.h.s.) describes the free motion of the particles with velocity $V\HAT n_1$ and the first term of the right-hand side (r.h.s.) accounts for the rotational diffusion of the director. $J$ is a term that accounts for the change in the distribution function due to collisions. To build $J$ an approach similar to the Boltzmann equation is used~\cite{chapman1990mathematical,soto2016kinetic}. That is, the  average collision rate is computed from the distribution function, and for each collision, the change in $f$ is obtained independently. Finally, $J$ is obtained by integrating over all possible collision parameters. 

A collision here is the encounter of two ABPs with directors $\HAT n_1$ and $\HAT n_2$, which are separated at the beginning of the collision by the vector $\sigma \hat{\bm{\sigma}}$ that goes from particle 1 to 2 (see Fig.~\ref{fig.appendix} in the Appendix). With this parametrization, the number of collisions per unit time is $\left.f^{(2)}(\VEC r_1, \HAT n_1,\VEC r_2,\HAT n_2)\right|_{|\VEC r_1-\VEC r_2|=\sigma} |\sigma V(\HAT n_2-\HAT n_1)\cdot\hat{\bm{\sigma}}|\Theta[-(\HAT n_2-\HAT n_1)\cdot\hat{\bm{\sigma}}]$, where $f^{(2)}$ is the two-particle distribution function that gives the number of pairs of particles having simultaneously positions and directors in the vicinity of $\VEC r_1$, $\HAT n_1$, $\VEC r_2$, and $\HAT n_2$. The factor $ |\sigma V(\HAT n_2-\HAT n_1)\cdot\hat{\bm{\sigma}}|$ takes into account that the collision rate is proportional to the particle size and the relative velocity normal to the relative vector. Finally,  the Heaviside step function $\Theta$ selects particles that are approaching. Using this collision rate would lead to a non-closed equation with the r.h.s. depending on $f^{(2)}$. To close the equation, we first write
\begin{align}
f^{(2)}(\VEC r_1, \HAT n_1,\VEC r_2,\HAT n_2) = g^{(2)}(\VEC r_1, \HAT n_1,\VEC r_2,\HAT n_2) f(\VEC r_1, \HAT n_1) f(\VEC r_2,\HAT n_2),
\end{align}
which defines the pair correlation function $g^{(2)}$. We recall that it must be evaluated only with the two particles in contact and approaching, that is, in a precollisional state, defining the pair correlation function at contact $\chi\equiv\lim_{r_{12}\to\sigma} \left.g^{(2)}(\VEC r_1, \HAT n_1,\VEC r_2,\HAT n_2)\right|_{(\HAT{n}_2-\HAT{n}_1)\cdot(\VEC r_2-\VEC r_1)<0}$. The classical molecular chaos hypothesis consists on assuming $\chi=1$. In dense gases of hard spheres particles it is common to go beyond this crude approximation and use the Enskog approach, which consists in assuming that $g^{(2)}$ at contact is given by the distribution at equilibrium, implying that $\chi$ depends only on the density profile~\cite{van1973modified,chapman1990mathematical,soto2016kinetic}. In granular gases, as a consequence of the longer times particles remain in contact as an effect of inelasticity, $g^{(2)}$ shows a large angular dependence for the postcollisional states, but for the precollisional states $\chi$ is very well approximated by the equilibrium expression for hard elastic spheres~\cite{brilliantov2004kinetic,garzo2019granular}. Also, there is a strong discontinuity at the transition between the pre- and post-collisional states, i.e. for grazing encounters~\cite{soto2001statistical}. 
Even for a simple homogeneous and isotropic state, the pair distribution function $g^{(2)}$ for ABPs depends in principle on the relative distance $r=|\VEC r_1 - \VEC r_2|$ and two angles. The measurements in Refs.~\onlinecite{de2018static,jeggle2020pair} show important angular dependencies,  but the authors did not separate pre- and postcollisional states, and therefore the measured values of $g^{(2)}$ cannot be used for a collision theory of ABPs. Therefore, in the absence of a validated expression, we will present the results in terms of a general $\chi$, but assuming that it depends only on the local density and not on the angles. 
 For explicit calculations, the elastic hard disk expression $\chi_\text{hd}=(1-7\eta/16)/(1-\eta)^2$ is used, where $\eta=\pi\sigma^2\rho_0/4$ is the area faction~\cite{henderson1975simple,PRL}.

When two particles meet, they slide until they reach a grazing configuration $(\HAT n_2-\HAT n_1)\cdot\hat{\bm{\sigma}}=0$, at which point they begin to move apart. In the limit $\Pe$ going to infinity, the  two particles kept their directors unchanged and it is direct to integrate the equations of motion of the two particles and obtain their final positions $\VEC r_{1,2}^\text{end}$ and the duration of the collision $\Delta t^\text{col}$. In absence of the collision, the final positions would have been the result of free motion, $\VEC r_{1,2}^\text{ini}+V\HAT{n}_{1,2}\Delta t^\text{col}$. Then the whole collision process can be interpreted as the particles being instantaneously displaced a distance $\VEC \Delta_{1,2}\equiv\VEC r_{1,2}^\text{end}-(\VEC r_{1,2}^\text{ini}+V\HAT{n}_{1,2}\Delta t^\text{col})$, while keeping the directors unchanged. Explicit expressions for the effective displacements $\VEC \Delta_{1,2}$ in terms of $\HAT{n}_{1,2}$ and $\hat{\bm{\sigma}}$, obtained in Ref.~\onlinecite{PRL}, are also given  in the Appendix. 

Considering all these elements, $J$ is written as a Boltzmann--Enskog-like collision term 
\begin{multline}\label{eq:operator}
    J[f] =  \int  \chi\left(\rho\left(\frac{\VEC r_1'+\VEC r_2'}{2}\right)\right) f(\VEC r_1',\HAT n_1) f(\VEC r_2',\HAT n_2)\\
%
\times |V\sigma^{d-1}(\HAT n_2-\HAT n_1)\cdot\hat{\bm{\sigma}}|\Theta[-(\HAT n_2-\HAT n_1)\cdot\hat{\bm{\sigma}}]
    \delta(\mathbf{r}_2^\prime-\mathbf{r}_1^\prime-\sigma\hat{\bm{\sigma}})\\
\times      [\delta(\mathbf{r}_1-\mathbf{r}_1^\prime-\VEC \Delta_1) -\delta(\mathbf{r}_1-\mathbf{r}_1^\prime)]d\mathbf{r}_1^\prime d\mathbf{r}_2^\prime d\mathbf{\hat{n}}_2d\hat{\bm{\sigma}}.
\end{multline}
It accounts for the effective displacement $\VEC \Delta_{1}$ of the tagged particle due to collisions with another particle of director $\HAT{n}_2$.
The loss term [associated with the Dirac delta $\delta(\mathbf{r}_1-\mathbf{r}_1^\prime)$] describes a particle at position $\mathbf{r}_1$ that collides with a partner, while the gain term, associated to the Dirac delta $\delta(\mathbf{r}_1-\mathbf{r}_1^\prime-\VEC \Delta_1)$, describes a particle that was at $\mathbf{r}_1-\VEC \Delta_1$ and ends  at $\mathbf{r}_1$  after the collision. In both cases, the partner is at a distance $\sigma\hat{\bm{\sigma}}$ from the tagged particle.
Finally, we recall that the collision term \eqref{eq:operator} describes the dynamics in the limit of very large $\Pe$, implying that the director $\HAT n_1$ of the tagged particle does not change during a collision.

\section{Kinetic equation for a tracer particle} \label{sec.kttagged}

To study the self-diffusion process, we consider a system of ABPs with number density $\rho_0$, in two spatial dimensions.  The system is at equilibrium, described by an isotropic and homogeneous distribution function $f_0=\rho_0/(2\pi)$. 
In the system, there is a tracer, a tagged particle, mechanically equal to the others. 
The motion of the tracer is described by the kinetic equation~\eqref{eq:kin}, where in the collision term the tagged particle  is described by an inhomogeneous and time-dependent distribution $f$, while the partners are described by the bath distribution $f_0$. Under these conditions, since $f_0$ does not depend on the position or the director, the collision term simplifies enormously and becomes a linear operator on $f$, called the Lorentz collision operator
\begin{multline}\label{eq:operatorL}
    J_\text{L}[f] = \frac{\rho_0\chi_0 \sigma}{2\pi} \int  \left[  f(\VEC r_1-\VEC \Delta_1,\HAT n_1)  - f(\VEC r_1,\HAT n_1)\right]   \\
\times |V(\HAT n_2-\HAT n_1)\cdot\hat{\bm{\sigma}}|\Theta[-(\HAT n_2-\HAT n_1)\cdot\hat{\bm{\sigma}}] d\mathbf{\hat{n}}_2d\hat{\bm{\sigma}}.
\end{multline}
With this, the evolution of $f$ for the tracer is given by the Boltzmann--Lorentz equation
\begin{equation}\label{eq:kinL}
    \frac{\partial f}{\partial t} + V\mathbf{\HAT{n}}_1\cdot\derpar{}{\VEC r_1} f = D_\text{r}\nabla^2_{\HAT{n}_1} f + J_\text{L}[f].
\end{equation}
The collision term can be easily interpreted as a loss and a gain term. For the loss term, the  tagged particle is collided by others causing it to be removed from its original position $\VEC r_1$. The gain term, on the other hand, describes a tagged particle that was located at $\VEC r_1-\VEC \Delta_1$, where the collision produces a net displacement $\VEC \Delta_1$, resulting in it ending up at $\VEC r_1$ after the collision. As in the full equation \eqref{eq:operator}, the factor $|V(\HAT n_2-\HAT n_1)\cdot\hat{\bm{\sigma}}|$ indicates that the collision rate is proportional to the relative velocity and $\Theta[-(\HAT n_2-\HAT n_1)\cdot\hat{\bm{\sigma}}]$ selects only particles that are approaching. 
Finally, for both processes the  collision frequency is proportional to $\rho_0\chi_0 \sigma$, where  $\chi_0$ is the pair correlation function at contact evaluated at the global density.  
In two spatial dimensions, directors are characterized by a single angle $\HAT n=(\cos\phi,\sin\phi)$ and the unit vector $\hat{\bm{\sigma}}$ is parametrized by the the angle $\theta$ it makes with $\HAT n_2 - \HAT n_1$, such that $\cos\theta<0$. More details on the parametrization of the integrals is given in the Appendix.

The resulting kinetic equation for the tracer \eqref{eq:operatorL} and \eqref{eq:kinL}, although linear, is complex to analyze under general conditions due to its intrinsic non-locality in space and the involved integrals in the angles. We will study the dynamics at times longer than the characteristic scale of collisions $t_\text{col}=\sigma/V$ and the mean free time $t_\text{free}=1/(\rho_0\chi_0V\sigma)$, where many individual collision events have taken place. Another relevant time scale is the  rotational diffusion time $t_\text{rot}=1/\Dr$, which characterizes the time for an ABP to forget its direction. Two cases will be studied, first the intermediate time scale $t_\text{col},t_\text{free}\ll t \ll t_\text{rot}$, where the tracer preserves is director but nevertheless it has experienced several collisions. Second, the long time regime, $t_\text{rot}\ll t$, is analyzed, in which case the tracer has  completely forgotten its original direction and the motion is purely diffusive. 
See Fig.~\ref{fig.figure} for a schematic representation of the different timescales.

To study the spatio-temporal evolution of $f$, we work with the Fourier--Laplace transform, which is defined for any function $A(\mathbf r,t)$ as 
\begin{align}
\tilde A(\mathbf k, s) =\int_0^\infty dt\, e^{-st}\int d^2r\, e^{-i\mathbf k\cdot\mathbf r} A(\mathbf r,t) .
\end{align}
Especially relevant is the Fourier--Laplace transform of the density
\begin{align}
\tilde\rho(\VEC k,s)=\int \tilde f(\mathbf k,\HAT{n},s) d\HAT{n},
\end{align}
also known as the van Hove function. For example, if the density is described by a simple diffusion equation in the small wavevector regime, then the van Hove function is given by $\tilde\rho(\VEC k,s)=1/(s+D k^2)$, which is a direct way to obtain the diffusion coefficient by simple inspection~\cite{boon1991molecular}. Note that the van Hove function is normalized to one, since there is a single tagged particle.

\section{Effective dynamics at intermediate time scales} \label{sec.intermediate}

To analyze the dynamics at the intermediate time scale, where rotational diffusion has not been effective in changing the tracer director, we take $\Dr=0$ in the kinetic equation and set as initial condition that the tracer is placed at the origin and has a well-defined direction, which is chosen to be $\phi=0$, meaning that $\HAT{n}=\HAT{x}$. That is,
\begin{align}
f(\mathbf r,\HAT{n},t=0)&=\delta(\mathbf r)\delta(\phi).
\end{align}
Suppressing  rotational diffusion eliminates the only mechanism for changing direction and therefore $\phi=0$ for all times, implying that the tracer distribution function can be written as 
\begin{align} \label{f0.mediumtime}
f(\mathbf r,\HAT{n},t)&=\rho(\mathbf r,t)\delta(\phi).
\end{align}
Here $\rho(\mathbf r,t)$ is the density, with the initial condition $\rho(\mathbf{r},t=0)=\delta(\mathbf r)$, which corresponds to the so-called van Hove process. Substituting Eq.~\eqref{f0.mediumtime} for $f$  in Eq.~\eqref{eq:kinL} and making a Laplace--Fourier transform of the equation gives
\begin{multline}
 s\tilde \rho - 1  + iV\mathbf k\cdot \mathbf{\HAT{x}} \tilde \rho = \Phi \tilde \rho \int   \left[e^{-i\mathbf k\cdot \VEC \Delta_1}-1\right] dX,
 \end{multline}
where $\Phi={\rho_0\chi V \sigma}/({2\pi})$,  $dX= |(\HAT n_2-\HAT n_1)\cdot\hat{\bm{\sigma}}|\Theta[-(\HAT n_2-\HAT n_1)\cdot\hat{\bm{\sigma}}]\big|_{\phi_1=0} d\phi_2 d\theta$, and the factor $\delta(\phi_1)$ was simplified in all terms. Note that in the third term in the l.h.s. we use that $\HAT{n}=\HAT{x}$ and therefore the streaming part  depends only on $k_x$.

The van Hove function is therefore given by
$\tilde \rho(\VEC k, s) = \left[s+ i V k_x + \Phi I  \right]^{-1}$, where 
 \begin{align}
I &=   \int   \left[1-e^{-i\mathbf k\cdot \VEC \Delta_1}\right]_{\phi_1=0} dX \nonumber\\
&= \int \left[i\mathbf k\cdot \VEC \Delta_1 + \frac{\VEC{k}\VEC{k}}{2}: \VEC \Delta_1\VEC \Delta_1+\dots \right]_{\phi_1=0} dX.
\end{align}
The collisional integral $I$ has been expanded in $k$ to analyze the long wavelength dynamics, stopping at order $k^2$, meaning that we analyze the streaming and diffusive behavior.
Since $\phi_1=0$ fixes a preferred direction, by symmetry arguments we have
\begin{subequations}
\begin{align}
\int \left. \VEC \Delta_1\right|_{\phi_1=0} dX &= A\HAT{x},\\
\frac{1}{2}\int \left.\VEC \Delta_1\VEC \Delta_1\right|_{\phi_1=0} dX &= B\HAT{x}\HAT{x}+C\HAT{y}\HAT{y}.
\end{align}
\end{subequations}
From these expressions we get
\begin{subequations}
\begin{align}
A&=\int \left. (\VEC \Delta_1\cdot\HAT{x})\right|_{\phi_1=0} dX = -\frac{\pi^2\sigma}{2} \approx -4.93\sigma,\\
B&=\frac{1}{2}\int \left.(\VEC \Delta_1\cdot\HAT{x})^2\right|_{\phi_1=0} dX = \frac{2(12G-1)\sigma^2}{9}\approx 2.22 \sigma^2,\\
C&=\frac{1}{2}\int \left.(\VEC \Delta_1\cdot\HAT{y})^2\right|_{\phi_1=0} dX = \frac{2(6G+1)\sigma^2}{9}\approx 1.44\sigma^2,
\end{align}
\end{subequations}
where $G\approx0.92$ is the Catalan's constant, and the integrals were computed using the methods described in the Appendix.

Collecting all terms and going back to a general fixed director $\HAT n$, the van Hove function is finally
\begin{align} \label{vanHove-intermediate}
\tilde \rho(\VEC k, s) = \left[s+ i \VEC{k}\cdot V_\text{eff}\HAT{n} + {\Phi B} (\VEC k\cdot\HAT{n})^2 + {\Phi C}(\VEC k\cdot\HAT{t})^2\right]^{-1},
\end{align}
where $\HAT{t}$ is the unit vector transversal to $\HAT{n}$ and 
\begin{align}
V_\text{eff}&=V+\Phi A=(1-\eta\chi_0)V \label{veff}
\end{align}
is the effective speed, which was previously found using an heuristic analysis of the kinetic equation~\cite{PRL}.  The decrease of the effective velocity with density is at the basis of the MIPS theory, where  expressions similar to Eq.~\eqref{veff} have been proposed based on semiempirical methods~\cite{bialke2013microscopic,solon2015active} or derived in the limiting case of infinite spatial dimensions~\cite{de2019active}. Here it emerged as an effect of the collisions without the need to impose it. By inspection, it is possible to identify the van Hove function \eqref{vanHove-intermediate} as the solution of the anisotropic advection-diffusion equation for the density
\begin{align}
\derpar{\rho}{t} +  V_\text{eff}\HAT{n}\cdot\nabla\rho = D_\parallel(\HAT n\cdot\nabla)^2\rho + D_\perp(\HAT t\cdot\nabla)^2\rho, \label{eq.rhointermediate1}
\end{align}
where $D_\parallel=\Phi B$ and $D_\perp=\Phi C$ are the longitudinal and transversal diffusion coefficients, respectively, which we note are different, being  the longitudinal one larger. With their density dependence through $\Phi$,
\begin{subequations}
\label{Dparperp}
\begin{align}
D_\parallel &= \frac{4(12G-1)\sigma^2}{9\pi^2} \eta\chi_0 V\sigma \approx 0.45 \eta\chi_0 V\sigma,\\
D_\perp &= \frac{4(6G+1)\sigma^2}{9\pi^2} \eta\chi_0 V\sigma \approx 0.29 \eta\chi_0 V\sigma.
\end{align}
\end{subequations}
Eq.~\eqref{eq.rhointermediate1} describes the effective motion of an ABP in the intermediate time scale, where the director remains fixed. Collisions cause its streaming velocity to decrease. But also, since collisions are stochastic, they generate diffusive motion with different diffusion coefficients in the longitudinal and transverse directions. A representation of this effective dynamics is shown in Fig.~\ref{fig.figure}c. 

Now, if we add back the rotational diffusion that takes place on a slower time scale, we obtain an effective Fokker--Planck equation for an ABP moving in a dense medium
\begin{align} \label{eq.FP}
\derpar{f}{t} +  V_\text{eff}\HAT{n}\cdot\nabla f = D_\parallel(\HAT n\cdot\nabla)^2f + D_\perp(\HAT t\cdot\nabla)^2f + D_\text{r}\nabla^2_{\HAT{n}} f,
\end{align}
where collisions are taken into account in the effective coefficients $V_\text{eff}$, $D_\parallel$, and $D_\perp$. 

On short time scales the motion is purely ballistic with the computed effective velocity, resulting in a mean square displacement  (MSD) that goes as $\langle \Delta r^2\rangle\sim V_\text{eff}^2 t^2$. Analogous results, modifying the MSD at short times due to interactions, have also been obtained for the self-diffusion process in single-file configurations~\cite{akintunde2024single}.

Finally, it is possible to extend this description to an ensemble of ABPs, where each particle follows the  stochastic dynamics associated with Eq.~\eqref{eq.FP}, but where now the coefficients depend on the local density field, representing mean-field interactions with the other particles. In this formalism, each particle is described by the Langevin equations
\begin{subequations}\label{eq.effectiveLangevin}
\begin{align} 
\dot {\VEC r}_i &=  V_\text{eff}(\eta_i)\HAT{n}_i + \sqrt{2 D_\parallel(\eta_i)} \xi_{\parallel i}\HAT n_i +\sqrt{2 D_\perp(\eta_i)}\xi_{\perp,i}\HAT t_i,\\
\dot{\phi}_i &= \sqrt{2\Dr}\xi_{\text{r},i},
\end{align}
\end{subequations}
where $\xi_{\parallel,i}$, $\xi_{\perp,i}$ and $\xi_{\text{r},i}$ uncorrelated white noises, and $\eta_i$ the coarse grained area fraction evaluated at $\VEC r_i$. The mean-field interaction makes Eqs.~\eqref{eq.effectiveLangevin}a-b significantly simpler than the original equations for ABPs interacting with hard-core repulsive pairwise potentials~\eqref{eq.ABP}, both for theoretical and numerical analysis.

Both effective diffusion coefficients \eqref{Dparperp} are proportional to $\chi_0$, which is expected to diverge at close packing, since collisions become infinitely frequent there. Therefore, both $D_\parallel$ and $D_\perp$ are predicted to  diverge as well. This result is evidently wrong because at close packing no motion takes place. The origin of this discrepancy is the fact that the kinetic equation assumes from the very beginning the existence of binary collisions, which is not the case at high densities where many-body effects are present. Furthermore, Eq.~\eqref{veff} shows that the theory cannot be valid beyond $\eta\chi_0=1$.  Indeed, beyond this density, the effective backward displacements due to collisions are larger than the free motion, resulting in a negative effective velocity. Using the elastic hard disk expression for the correlation function, this condition gives $\eta_\text{max}\approx0.42$.


\section{Long-time diffusivity} \label{sec.long}
For times greater than $D_r$, the tracer has  completely forgotten its original direction and moves in an effective random walk. For isolated particles, this long-time diffusivity can first be estimated  by dimensional analysis by noting that the particle performs a correlated motion with lengths $\ell=V/\Dr$ for times lasting $\tau=\Dr^{-1}$. This results in a diffusion coefficient that scales as $\ell^2/\tau=V^2/\Dr$. Precise analytical calculations in Ref.~\onlinecite{howse2007self} for isolated ABPs give the bare diffusivity $D_0=V^2/(2\Dr)$. In the presence of other particles, the velocity is effectively reduced as shown in the previous section, but collisional diffusive motion also appears. 
To obtain the long-time diffusion coefficient from the Lorentz equation \eqref{eq:kinL}, we  again consider the van Hove process: the tracer is initially placed at the origin, but now with an isotropic distribution
\begin{align}
f(\mathbf r,\HAT{n},t=0)&=\delta(\mathbf r)/(2\pi),
\end{align}
which is associated with an initial density $\rho(\mathbf r,t=0)=\delta(\mathbf r)$. Figure~\ref{fig.figure}d represents  this process schematically.
In this regime, as the director evolves in time, it is not possible to obtain a closed expression for the van Hove function as in the previous section. Nevertheless, as we will show below, it is possible to compute the diffusion coefficient analytically with no approximation. To do this, we take advantage of the fact that the Laplace transform of the MSD can be computed in 2D as shown in Refs.~\onlinecite{boon1991molecular,villa2020run},
\begin{align}\label{eq.MSDs}
\langle\widetilde{\Delta r^2}\rangle(s) &= \lim_{k\to 0}\left( -2 \derpar{^2\tilde \rho}{k^2}\right).
\end{align}
At long times, the MSD should grow linearly with time $\langle{\Delta r^2}\rangle(t) \sim 4 D t$, reflecting a diffusion process with  diffusion coefficient $D$.
Recalling that the Laplace transform of $t$ is $1/s^2$, the diffusion coefficient can, therefore, be obtained as the $s^{-2}$ coefficient of the Laurent series of the MSD in Laplace space
 \begin{align} \label{eq.DfromMSD}
\langle\widetilde{\Delta r^2}\rangle(s) &= \frac{4 D}{s^2} + \sum_{n\geq -1} a_n s^n.
\end{align}

The objective then is to solve the kinetic equation in Fourier--Laplace space. From this, the procedure is to compute the van Hove function, then to obtain the MSD in Laplace space, and finally to derive the diffusion coefficient. 
Using the initial condition, the kinetic equation~\eqref{eq:kinL} for $\tilde f(\mathbf k,\HAT n_1,s)$, reads
\begin{align}\label{eq:kin-ks}
    s\tilde f - \frac{1}{2\pi}  + iV\mathbf k\cdot \mathbf{\HAT{n}}_1 \tilde f - D_\text{r}\derpar{^2\tilde f}{\phi_1^2} = \Phi \int  \left[ e^{-i\mathbf k\cdot \VEC \Delta_1}-1\right]\tilde f  dX.
\end{align}
To solve it, we use an expansion in Fourier angular modes
\begin{align}
\tilde f(\VEC k,\HAT n_1,s) =\frac{1}{2\pi}\sum_m g_m(\VEC k,s) e^{i m \phi_1},
\end{align}
where the factor $1/2\pi$ is written  for convenience, so that the van Hove function is  $\tilde\rho(\mathbf k, s) = g_0(\mathbf k, s)$.
Projecting the kinetic equation into the $p$ mode and using $\VEC k=k\HAT x$, we get
\begin{align}\label{eq.gp}
s g_p  +\frac{i kV}{2}(g_{p+1}+g_{p-1}) +  D_\text{r} p^2 g_p + \Phi  \sum_m L_{pm} g_m &= \delta_{p,0},
\end{align}
which can be written as $\sum_m A_{pm} g_m = \delta_{p,0}$, defining the matrix $\mathbb{A}$. This gives $\tilde\rho(\mathbf k, s)=g_0=[\mathbb{A}^{-1}]_{00}$. Here, 
\begin{multline}\label{q.defIpm}
L_{pm} =\frac{1}{2\pi} \int  e^{i(m-p)\phi_1} \left[1-e^{-i\mathbf k\cdot \VEC \Delta_1}\right]\\
\times |(\HAT n_2-\HAT n_1)\cdot\hat{\bm{\sigma}}|\Theta[-(\HAT n_2-\HAT n_1)\cdot\hat{\bm{\sigma}}] d\phi_1 d\phi_2 d\theta ,
\end{multline}
which depend only on $m-p$. 
 First, without  explicitly computing the matrix elements $L_{pm}$,  it is possible using symbolic algebra softwares to obtain $\tilde\rho(\mathbf k, s)$ for increasing number of angular Fourier modes $P$, with expressions that depend on $L_{pm}$. Since we aim to compute the MSD via Eq.~\eqref{eq.MSDs},  we make a Taylor expansion in $k$ of $L_{pm}$, by simply expanding the square bracket in the integral,
which gives $L_{pm}(k) = L^{(1)}_{pm} k +  L^{(2)}_{pm} k^2 + L^{(3)}_{pm} k^3 +\dots$, where it is directly verified by inspection that the constant terms vanish. This procedure gives for each $P$ the MSD in Laplace space together with its Laurent series in $s$. The leading term goes as  $L_{00}^{(1)}/s^3$, which would imply a superdiffusive $t^2$ behavior. Using the parametrization given in the Appendix, it is direct to compute the integrals and obtain that  $L_{00}^{(1)}$ vanishes identically. In fact, this term is proportional to the collisional average of $\mathbf{\Delta}_1$, which vanishes by isotropy. This results in  the leading order in the MSD going, as expected, as $1/s^2$, which allows us to obtain the diffusion coefficient using Eq.~\eqref{eq.DfromMSD}
\begin{multline}
D = \frac{V^2}{2\Dr} + \Phi \bigg[ L_{00}^{(2)} - \frac{iV}{2\Dr}\left(L_{-1,0}^{(1)} + L_{0,-1}^{(1)} + L_{1,0}^{(1)} + L_{0,1}^{(1)} \right)\\
 - \sum_{m=1}^{\infty} \frac{L_{-m,0}^{(1)} L_{0,-m}^{(1)} + L_{m,0}^{(1)} L_{0,m}^{(1)} }{m^2 \Dr} \bigg], \label{eq.Dserie}
\end{multline}
where we recognize the first term as the bare diffusivity $D_0$ for collisionless ABPs, while the second term results from the effect of collisions as it is manifest by the presence of the prefactor $\Phi$, which is proportional to the cross section and the pair correlation function. 
Thus, to evaluate $D$, one needs to compute only a few matrix elements of order $k^1$, and only $L_{00}$ of order $k^2$. Using the parametrization described in the Appendix, it is obtained that
$L_{p,p\pm1}^{(1)} = - i\pi^2\sigma/4$ and zero for the rest, which means that the sum in Eq.~\eqref{eq.Dserie} stops at $m=1$. Also, $L_{00}^{(2)} = 2 G\sigma^2$.
 Collecting all terms gives
\begin{align}
D &=  \frac{\left[V(1-\rho_0\chi_0\pi^2  \sigma/4)\right]^2}{2\Dr} + G\rho_0\chi_0 V \sigma^3/\pi.
\end{align}
The first term turns out to be simply the active diffusivity but now with the effective velocity $V_\text{eff}$ given in Eq.~\eqref{veff}. In dimensionless form
$D =  D_0 (1-\chi_0\eta)^2 + D_0 \frac{8 G \chi_0\eta}{\pi^2{\rm Pe}}$.
Since the kinetic theory we started with is  valid only for high Péclet number, it is not consistent to consider the second term, and therefore the prediction of the theory reduces to
\begin{align} \label{eq.Dlongterm}
D &= \frac{V_\text{eff}^2}{2\Dr} = D_0 (1-\chi_0\eta)^2.
\end{align}
Notably, this dependence on the effective velocity has  already been observed in numerical simulations of ABPs~\cite{wang2019anomalous}. 
As in the case of the dynamics at intermediate time scales, here the effective diffusion coefficient also depends on $\chi_0$. Also the theory fails at $\chi_0\eta=1$, where $D$ vanishes, but most likely the theory becomes quantitatively inaccurate for smaller densities. 

Using the pair correlation function of elastic disks, it is possible to compute explicitly the long-term diffusivity for various values of the area fraction: $D(\eta=0.1)\approx 0.78D_0$,  $D(\eta=0.2)\approx 0.51 D_0$, and  $D(\eta=0.3)\approx 0.22D_0$, and  $D(\eta=0.4)\approx 0.007D_0$. These values  can be compared with the numerical simulations in Ref.~\onlinecite{wang2019anomalous}:  $D(\eta=0.1)\approx 0.82D_0$,  $D(\eta=0.2)\approx 0.64D_0$, and  $D(\eta=0.3)\approx 0.46D_0$, and  $D(\eta=0.4)\approx 0.28D_0$. This discrepancy, with computed values smaller than the measured ones suggest that the use of the equilibrium expression for the pair correlation function is a too crude approximation and it overestimates its real value.

\section{Conclusions} \label{sec.conclusions}

A recently proposed kinetic theory for interacting ABPs is used to study their self-diffusive dynamics.
In the kinetic equation, the persistent excluded volume interactions between particles are described as effective particle displacements. To calculate the displacements, the kinetic theory assumes that the P\'eclet number is large, which implies that the directors of the particles do not change during collisions.

Starting from this kinetic equation, we consider the dynamics of a tagged particle, identical to  the ABP bath, which is in an equilibrium distribution. On intermediate time scales, where the director of the tagged particle remains fixed, the effective displacements produce an  anisotropic diffusion, which is added to the streaming motion with an  reduced effective velocity. This is summarized by the Fokker--Planck equation \eqref{eq.FP}. Extending this description to an ensemble of interacting particles, this effective dynamics can be mapped to a set of Langevin equations where  the particles now interact  in a mean-field manner through the local density [Eqs.~\eqref{eq.effectiveLangevin}a-b].

On larger time scales, where the director has had time to change, the dynamics is purely diffusive, with a diffusion coefficient given by Eq.~\eqref{eq.Dlongterm} that accounts for all interaction effects. Notably, the long term diffusivity can be written as that of free particles, but now with the  effective velocity.

The effective velocity  becomes negative at large densities, which is clearly unphysical and affects the prediction on both intermediate and large timescales. Since the effective velocity was obtained without approximation from the kinetic theory, this implies that it is the kinetic theory that is limited to moderate densities. The vanishing of the effective velocity marks a strong maximum density for the theory to be valid, which using the pair correlation function of elastic hard disks gives $\eta\approx 0.42$. On the other hand, in Ref.~\onlinecite{PRL} it was shown that the kinetic theory correctly predicts that the spinodal density for MIPS is $\eta\approx0.25$ in two dimensions. These results imply that the present description is limited to moderate densities, quantitatively valid up to $\eta\approx0.25$, to gradually deteriorate for increasing densities and to break down completely at $\eta\approx 0.42$. 
For area fractions higher than 0.25, the system will spontaneously separate into dilute and dense phases due to MIPS, and it is expected that the present calculations of self-diffusive motion  remain valid as long as the density of the dense phase is not too high. Also, the numerical comparison of the long-term diffusivity with numerical simulations indicate that the equilibrium expression for the pair correlation function overestimates its value for ABPs.
Further research is therefore needed to include, at least phenomenologically, many-body effects and the suppression of motion at high densities. This includes the study of the pair correlation function at contact for precollisional states and its possible angular dependence. If this is the case, and $\chi$ depends on the angles, the collisional integrals can be carried out according to the methods described above, modifying only the numerical values of the coefficients, which will depend on the explicit dependencies of $\chi$. 

In Ref.~\onlinecite{dulaney2021isothermal} it is shown that, similarly to equilibrium systems, for ABPs the long-time self-diffusivity $D$ can be expressed in terms of the static structure factor and the isothermal compressibility. An interesting avenue of research is to investigate whether these relations remain valid within the framework of kinetic theory. Both density fluctuations and a protocol for calculating stresses in kinetic theory should be worked out, which are left for future work.

Finally, the results presented in this article are based on a kinetic theory that is limited to high P\'eclet numbers, and  are therefore quantitatively correct only in this regime. The extension of the theory to finite P\'eclet needs to describe collisions stochastically, since the moment when the collision ends is a random variable. Ongoing research in this direction is on course and will be published elsewhere.


\acknowledgments
This research was supported by the Fondecyt Grant No.~1220536 and Millennium Science Initiative Program NCN19\_170 of ANID, Chile. R.S.\ thanks Ricardo Brito for inspiring discussions.

\appendix
\section{Effective displacements and collision parametrization in two spatial dimensions}

\begin{figure}[htb]
\includegraphics[width=\columnwidth]{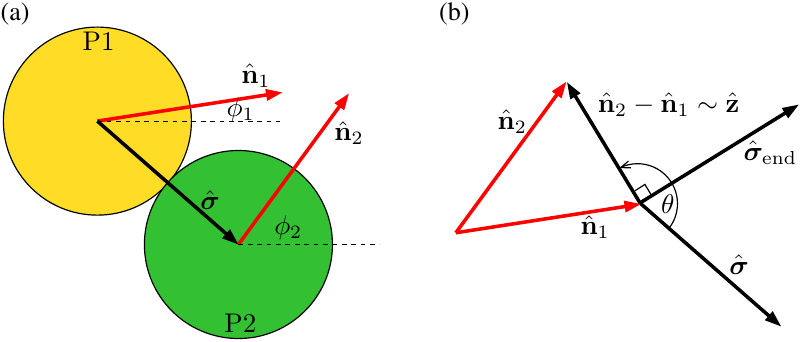}
\caption{Geometry of the collision or particles P1 and P2. (a) The particle directors $\HAT n_1$ and $\HAT n_2$ are parametrized by the angles $\phi_{1,2}$, respectively, measured with respect to a fixed horizontal axis. The unit vector $\hat{\bm{\sigma}}$ points from particle 1 to particle 2 at the beginning of the collision. (b) $\hat{\bm{\sigma}}$ is parametrized with the angle $\theta$, which it forms with $\HAT{z}=(\HAT{n}_2-\HAT{n}_1)/|\HAT{n}_2-\HAT{n}_1|$. At the end of the collision, the unit vector from particle 1 to particle 2 is $\hat{\bm{\sigma}}_\text{end}$.}
\label{fig.appendix}
\end{figure}

Here, for completeness, we copy from Ref.~\onlinecite{PRL} the expressions for the effective displacements and the form the collisional integrals are computed. In two dimensions, collisions are parametrized by the angles $\phi_1$ and $\phi_2$ of the two directors, and $\theta$ the angle formed by the unit vector $\hat{\bm{\sigma}}$ and  $\HAT n_2 - \HAT n_1$ at the beginning of the collision, such that $\cos\theta<0$, as shown in Fig.~\ref{fig.appendix}. 

Integrating the equations of motion it results that the duration of the collision is~\cite{PRL}
\begin{align}
\Delta t^\text{col}= \frac{\sigma}{V|\HAT{n}_2-\HAT{n}_1|} \log|\tan(\theta/2)|,
\end{align}
and two situations can can take place for the angle $\theta_\text{end}$ at the end of the collision. If $\pi/2\leq\theta<\pi$,  $\theta_\text{end}=\pi/2$, but if $\pi<\theta\leq3\pi/2$, $\theta_\text{end}=3\pi/2$. With this, writing the directors as $\HAT{n}_i=(\cos\phi_i,\sin\phi_i)$, the unit vectors can be written as
\begin{align}
\hat{\bm{\sigma}}&=R(-\theta) \HAT{z},\\
\intertext{and}
\hat{\bm{\sigma}}_\text{end}&=\begin{cases}
R(-\pi/2) \HAT{z}, & \pi/2\leq\theta<\pi\\
R(\pi/2) \HAT{z}, & \pi<\theta\leq3\pi/2,
\end{cases}
\end{align}
where $\HAT{z} = \left(\frac{\cos\phi_2-\cos\phi_1}{\sqrt{2-2\cos(\phi_2-\phi_1)}}, \frac{\sin\phi_2-\sin\phi_1}{\sqrt{2-2\cos(\phi_2-\phi_1)}}\right)$ is the unit vector parallel to $(\HAT n_2 - \HAT n_1)$ and 
$R(\alpha)=\begin{pmatrix}
\cos\alpha & -\sin\alpha\\
\sin\alpha & \cos\alpha
\end{pmatrix}$ is the rotation matrix.
Finally, the effective displacements are
\begin{align}
\VEC{\Delta}_1 &= -\sigma\frac{\hat{\bm{\sigma}}_\text{end}-\hat{\bm{\sigma}}_0}{2}
- V\Delta t^\text{col} \frac{\HAT{n}_1-\HAT{n}_2}{2},\\
\VEC{\Delta}_2 &= -\VEC{\Delta}_1.
\end{align}
With this parametrization, the different expressions needed to compute the integrals can be easily expressed in terms of $\phi_1$, $\phi_2$, and $\theta$. For example,
\begin{gather}
|\HAT{n}_2-\HAT{n}_1| = \sqrt{2 - 2 \cos(\phi_1 - \phi_2)},\\
(\HAT n_2-\HAT n_1)\cdot\hat{\bm{\sigma}}= \sqrt{2 - 2 \cos(\phi_1 - \phi_2)}\cos\theta,\\
    |(\HAT n_2-\HAT n_1)\cdot\hat{\bm{\sigma}} | \bm{\Delta}_1=\frac{\sigma|\cos\theta|}{2} \bigg\{\big[\cos(\phi_2 - \theta) - \nonumber \\
   \cos(\phi_1 - \theta)    + (\cos\phi_2 - \cos\phi_1) \log
     |\tan(\theta/2)| \nonumber \\
     \pm (\sin\phi_1 - \sin\phi_2)\big]\HAT{x} 
+ \big[\sin(\phi_2 - \theta) - 
   \sin(\phi_1 - \theta) \nonumber \\+ (\sin\phi_2 - \sin\phi_1) \log
     |\tan(\theta/2)| 
     \pm (\cos\phi_2 - \cos\phi_1)\big]\HAT{y}\bigg\},\label{SM.Deltaexplicit}
\end{gather}
where the positive sign is for $\pi/2\leq\theta<\pi$ and the negative one for $\pi<\theta\leq3\pi/2$.


\begin{thebibliography}{38}%
\makeatletter
\providecommand \@ifxundefined [1]{%
 \@ifx{#1\undefined}
}%
\providecommand \@ifnum [1]{%
 \ifnum #1\expandafter \@firstoftwo
 \else \expandafter \@secondoftwo
 \fi
}%
\providecommand \@ifx [1]{%
 \ifx #1\expandafter \@firstoftwo
 \else \expandafter \@secondoftwo
 \fi
}%
\providecommand \natexlab [1]{#1}%
\providecommand \enquote  [1]{``#1''}%
\providecommand \bibnamefont  [1]{#1}%
\providecommand \bibfnamefont [1]{#1}%
\providecommand \citenamefont [1]{#1}%
\providecommand \href@noop [0]{\@secondoftwo}%
\providecommand \href [0]{\begingroup \@sanitize@url \@href}%
\providecommand \@href[1]{\@@startlink{#1}\@@href}%
\providecommand \@@href[1]{\endgroup#1\@@endlink}%
\providecommand \@sanitize@url [0]{\catcode `\\12\catcode `\$12\catcode
  `\&12\catcode `\#12\catcode `\^12\catcode `\_12\catcode `\%12\relax}%
\providecommand \@@startlink[1]{}%
\providecommand \@@endlink[0]{}%
\providecommand \url  [0]{\begingroup\@sanitize@url \@url }%
\providecommand \@url [1]{\endgroup\@href {#1}{\urlprefix }}%
\providecommand \urlprefix  [0]{URL }%
\providecommand \Eprint [0]{\href }%
\providecommand \doibase [0]{http://dx.doi.org/}%
\providecommand \selectlanguage [0]{\@gobble}%
\providecommand \bibinfo  [0]{\@secondoftwo}%
\providecommand \bibfield  [0]{\@secondoftwo}%
\providecommand \translation [1]{[#1]}%
\providecommand \BibitemOpen [0]{}%
\providecommand \bibitemStop [0]{}%
\providecommand \bibitemNoStop [0]{.\EOS\space}%
\providecommand \EOS [0]{\spacefactor3000\relax}%
\providecommand \BibitemShut  [1]{\csname bibitem#1\endcsname}%
\let\auto@bib@innerbib\@empty
\bibitem [{\citenamefont {Bechinger}\ \emph {et~al.}(2016)\citenamefont
  {Bechinger}, \citenamefont {Di~Leonardo}, \citenamefont {L{\"o}wen},
  \citenamefont {Reichhardt}, \citenamefont {Volpe},\ and\ \citenamefont
  {Volpe}}]{bechinger2016active}%
  \BibitemOpen
  \bibfield  {author} {\bibinfo {author} {\bibfnamefont {C.}~\bibnamefont
  {Bechinger}}, \bibinfo {author} {\bibfnamefont {R.}~\bibnamefont
  {Di~Leonardo}}, \bibinfo {author} {\bibfnamefont {H.}~\bibnamefont
  {L{\"o}wen}}, \bibinfo {author} {\bibfnamefont {C.}~\bibnamefont
  {Reichhardt}}, \bibinfo {author} {\bibfnamefont {G.}~\bibnamefont {Volpe}}, \
  and\ \bibinfo {author} {\bibfnamefont {G.}~\bibnamefont {Volpe}},\ }\bibfield
   {title} {\enquote {\bibinfo {title} {Active particles in complex and crowded
  environments},}\ }\href@noop {} {\bibfield  {journal} {\bibinfo  {journal}
  {Reviews of modern physics}\ }\textbf {\bibinfo {volume} {88}},\ \bibinfo
  {pages} {045006} (\bibinfo {year} {2016})}\BibitemShut {NoStop}%
\bibitem [{\citenamefont {Berg}(1993)}]{berg1993random}%
  \BibitemOpen
  \bibfield  {author} {\bibinfo {author} {\bibfnamefont {H.~C.}\ \bibnamefont
  {Berg}},\ }\href@noop {} {\emph {\bibinfo {title} {Random walks in
  biology}}}\ (\bibinfo  {publisher} {Princeton University Press},\ \bibinfo
  {year} {1993})\BibitemShut {NoStop}%
\bibitem [{\citenamefont {Berg}(2004)}]{Berg2001E.ColiMotion}%
  \BibitemOpen
  \bibfield  {author} {\bibinfo {author} {\bibfnamefont {H.~C.}\ \bibnamefont
  {Berg}},\ }\href@noop {} {\emph {\bibinfo {title} {E. coli in Motion}}}\
  (\bibinfo  {publisher} {Springer},\ \bibinfo {year} {2004})\BibitemShut
  {NoStop}%
\bibitem [{\citenamefont {Stramer}\ and\ \citenamefont
  {Mayor}(2017)}]{stramer2017mechanisms}%
  \BibitemOpen
  \bibfield  {author} {\bibinfo {author} {\bibfnamefont {B.}~\bibnamefont
  {Stramer}}\ and\ \bibinfo {author} {\bibfnamefont {R.}~\bibnamefont
  {Mayor}},\ }\bibfield  {title} {\enquote {\bibinfo {title} {Mechanisms and in
  vivo functions of contact inhibition of locomotion},}\ }\href@noop {}
  {\bibfield  {journal} {\bibinfo  {journal} {Nature reviews Molecular cell
  biology}\ }\textbf {\bibinfo {volume} {18}},\ \bibinfo {pages} {43--55}
  (\bibinfo {year} {2017})}\BibitemShut {NoStop}%
\bibitem [{\citenamefont {Paxton}\ \emph {et~al.}(2004)\citenamefont {Paxton},
  \citenamefont {Kistler}, \citenamefont {Olmeda}, \citenamefont {Sen},
  \citenamefont {St.~Angelo}, \citenamefont {Cao}, \citenamefont {Mallouk},
  \citenamefont {Lammert},\ and\ \citenamefont {Crespi}}]{paxton2004catalytic}%
  \BibitemOpen
  \bibfield  {author} {\bibinfo {author} {\bibfnamefont {W.~F.}\ \bibnamefont
  {Paxton}}, \bibinfo {author} {\bibfnamefont {K.~C.}\ \bibnamefont {Kistler}},
  \bibinfo {author} {\bibfnamefont {C.~C.}\ \bibnamefont {Olmeda}}, \bibinfo
  {author} {\bibfnamefont {A.}~\bibnamefont {Sen}}, \bibinfo {author}
  {\bibfnamefont {S.~K.}\ \bibnamefont {St.~Angelo}}, \bibinfo {author}
  {\bibfnamefont {Y.}~\bibnamefont {Cao}}, \bibinfo {author} {\bibfnamefont
  {T.~E.}\ \bibnamefont {Mallouk}}, \bibinfo {author} {\bibfnamefont {P.~E.}\
  \bibnamefont {Lammert}}, \ and\ \bibinfo {author} {\bibfnamefont {V.~H.}\
  \bibnamefont {Crespi}},\ }\bibfield  {title} {\enquote {\bibinfo {title}
  {Catalytic nanomotors: autonomous movement of striped nanorods},}\
  }\href@noop {} {\bibfield  {journal} {\bibinfo  {journal} {Journal of the
  American Chemical Society}\ }\textbf {\bibinfo {volume} {126}},\ \bibinfo
  {pages} {13424--13431} (\bibinfo {year} {2004})}\BibitemShut {NoStop}%
\bibitem [{\citenamefont {Bricard}\ \emph {et~al.}(2013)\citenamefont
  {Bricard}, \citenamefont {Caussin}, \citenamefont {Desreumaux}, \citenamefont
  {Dauchot},\ and\ \citenamefont {Bartolo}}]{bricard2013emergence}%
  \BibitemOpen
  \bibfield  {author} {\bibinfo {author} {\bibfnamefont {A.}~\bibnamefont
  {Bricard}}, \bibinfo {author} {\bibfnamefont {J.-B.}\ \bibnamefont
  {Caussin}}, \bibinfo {author} {\bibfnamefont {N.}~\bibnamefont {Desreumaux}},
  \bibinfo {author} {\bibfnamefont {O.}~\bibnamefont {Dauchot}}, \ and\
  \bibinfo {author} {\bibfnamefont {D.}~\bibnamefont {Bartolo}},\ }\bibfield
  {title} {\enquote {\bibinfo {title} {Emergence of macroscopic directed motion
  in populations of motile colloids},}\ }\href@noop {} {\bibfield  {journal}
  {\bibinfo  {journal} {Nature}\ }\textbf {\bibinfo {volume} {503}},\ \bibinfo
  {pages} {95--98} (\bibinfo {year} {2013})}\BibitemShut {NoStop}%
\bibitem [{\citenamefont {Villa-Torrealba}\ \emph {et~al.}(2020)\citenamefont
  {Villa-Torrealba}, \citenamefont {Ch{\'a}vez-Raby}, \citenamefont
  {de~Castro},\ and\ \citenamefont {Soto}}]{villa2020run}%
  \BibitemOpen
  \bibfield  {author} {\bibinfo {author} {\bibfnamefont {A.}~\bibnamefont
  {Villa-Torrealba}}, \bibinfo {author} {\bibfnamefont {C.}~\bibnamefont
  {Ch{\'a}vez-Raby}}, \bibinfo {author} {\bibfnamefont {P.}~\bibnamefont
  {de~Castro}}, \ and\ \bibinfo {author} {\bibfnamefont {R.}~\bibnamefont
  {Soto}},\ }\bibfield  {title} {\enquote {\bibinfo {title} {Run-and-tumble
  bacteria slowly approaching the diffusive regime},}\ }\href@noop {}
  {\bibfield  {journal} {\bibinfo  {journal} {Physical Review E}\ }\textbf
  {\bibinfo {volume} {101}},\ \bibinfo {pages} {062607} (\bibinfo {year}
  {2020})}\BibitemShut {NoStop}%
\bibitem [{\citenamefont {Howse}\ \emph {et~al.}(2007)\citenamefont {Howse},
  \citenamefont {Jones}, \citenamefont {Ryan}, \citenamefont {Gough},
  \citenamefont {Vafabakhsh},\ and\ \citenamefont
  {Golestanian}}]{howse2007self}%
  \BibitemOpen
  \bibfield  {author} {\bibinfo {author} {\bibfnamefont {J.~R.}\ \bibnamefont
  {Howse}}, \bibinfo {author} {\bibfnamefont {R.~A.}\ \bibnamefont {Jones}},
  \bibinfo {author} {\bibfnamefont {A.~J.}\ \bibnamefont {Ryan}}, \bibinfo
  {author} {\bibfnamefont {T.}~\bibnamefont {Gough}}, \bibinfo {author}
  {\bibfnamefont {R.}~\bibnamefont {Vafabakhsh}}, \ and\ \bibinfo {author}
  {\bibfnamefont {R.}~\bibnamefont {Golestanian}},\ }\bibfield  {title}
  {\enquote {\bibinfo {title} {Self-motile colloidal particles: from directed
  propulsion to random walk},}\ }\href@noop {} {\bibfield  {journal} {\bibinfo
  {journal} {Physical review letters}\ }\textbf {\bibinfo {volume} {99}},\
  \bibinfo {pages} {048102} (\bibinfo {year} {2007})}\BibitemShut {NoStop}%
\bibitem [{\citenamefont {Lindner}\ and\ \citenamefont
  {Nicola}(2008)}]{lindner2008diffusion}%
  \BibitemOpen
  \bibfield  {author} {\bibinfo {author} {\bibfnamefont {B.}~\bibnamefont
  {Lindner}}\ and\ \bibinfo {author} {\bibfnamefont {E.}~\bibnamefont
  {Nicola}},\ }\bibfield  {title} {\enquote {\bibinfo {title} {Diffusion in
  different models of active brownian motion},}\ }\href@noop {} {\bibfield
  {journal} {\bibinfo  {journal} {The European Physical Journal Special
  Topics}\ }\textbf {\bibinfo {volume} {157}},\ \bibinfo {pages} {43--52}
  (\bibinfo {year} {2008})}\BibitemShut {NoStop}%
\bibitem [{\citenamefont {Romanczuk}\ \emph {et~al.}(2012)\citenamefont
  {Romanczuk}, \citenamefont {B{\"a}r}, \citenamefont {Ebeling}, \citenamefont
  {Lindner},\ and\ \citenamefont {Schimansky-Geier}}]{romanczuk2012active}%
  \BibitemOpen
  \bibfield  {author} {\bibinfo {author} {\bibfnamefont {P.}~\bibnamefont
  {Romanczuk}}, \bibinfo {author} {\bibfnamefont {M.}~\bibnamefont {B{\"a}r}},
  \bibinfo {author} {\bibfnamefont {W.}~\bibnamefont {Ebeling}}, \bibinfo
  {author} {\bibfnamefont {B.}~\bibnamefont {Lindner}}, \ and\ \bibinfo
  {author} {\bibfnamefont {L.}~\bibnamefont {Schimansky-Geier}},\ }\bibfield
  {title} {\enquote {\bibinfo {title} {Active brownian particles: From
  individual to collective stochastic dynamics},}\ }\href@noop {} {\bibfield
  {journal} {\bibinfo  {journal} {The European Physical Journal Special
  Topics}\ }\textbf {\bibinfo {volume} {202}},\ \bibinfo {pages} {1--162}
  (\bibinfo {year} {2012})}\BibitemShut {NoStop}%
\bibitem [{\citenamefont {Kumar}\ \emph {et~al.}(2021)\citenamefont {Kumar},
  \citenamefont {Singh}, \citenamefont {Giri},\ and\ \citenamefont
  {Mishra}}]{kumar2021effect}%
  \BibitemOpen
  \bibfield  {author} {\bibinfo {author} {\bibfnamefont {S.}~\bibnamefont
  {Kumar}}, \bibinfo {author} {\bibfnamefont {J.~P.}\ \bibnamefont {Singh}},
  \bibinfo {author} {\bibfnamefont {D.}~\bibnamefont {Giri}}, \ and\ \bibinfo
  {author} {\bibfnamefont {S.}~\bibnamefont {Mishra}},\ }\bibfield  {title}
  {\enquote {\bibinfo {title} {Effect of polydispersity on the dynamics of
  active brownian particles},}\ }\href@noop {} {\bibfield  {journal} {\bibinfo
  {journal} {Physical Review E}\ }\textbf {\bibinfo {volume} {104}},\ \bibinfo
  {pages} {024601} (\bibinfo {year} {2021})}\BibitemShut {NoStop}%
\bibitem [{\citenamefont {Szamel}\ and\ \citenamefont
  {Flenner}(2024)}]{szamel2024extremely}%
  \BibitemOpen
  \bibfield  {author} {\bibinfo {author} {\bibfnamefont {G.}~\bibnamefont
  {Szamel}}\ and\ \bibinfo {author} {\bibfnamefont {E.}~\bibnamefont
  {Flenner}},\ }\bibfield  {title} {\enquote {\bibinfo {title} {Extremely
  persistent dense active fluids},}\ }\href@noop {} {\bibfield  {journal}
  {\bibinfo  {journal} {Soft Matter}\ }\textbf {\bibinfo {volume} {20}},\
  \bibinfo {pages} {5237--5244} (\bibinfo {year} {2024})}\BibitemShut {NoStop}%
\bibitem [{\citenamefont {Cates}\ and\ \citenamefont
  {Tailleur}(2013)}]{cates2013active}%
  \BibitemOpen
  \bibfield  {author} {\bibinfo {author} {\bibfnamefont {M.~E.}\ \bibnamefont
  {Cates}}\ and\ \bibinfo {author} {\bibfnamefont {J.}~\bibnamefont
  {Tailleur}},\ }\bibfield  {title} {\enquote {\bibinfo {title} {When are
  active brownian particles and run-and-tumble particles equivalent?
  consequences for motility-induced phase separation},}\ }\href@noop {}
  {\bibfield  {journal} {\bibinfo  {journal} {Europhysics Letters}\ }\textbf
  {\bibinfo {volume} {101}},\ \bibinfo {pages} {20010} (\bibinfo {year}
  {2013})}\BibitemShut {NoStop}%
\bibitem [{\citenamefont {Solon}, \citenamefont {Cates},\ and\ \citenamefont
  {Tailleur}(2015)}]{solon2015active}%
  \BibitemOpen
  \bibfield  {author} {\bibinfo {author} {\bibfnamefont {A.~P.}\ \bibnamefont
  {Solon}}, \bibinfo {author} {\bibfnamefont {M.~E.}\ \bibnamefont {Cates}}, \
  and\ \bibinfo {author} {\bibfnamefont {J.}~\bibnamefont {Tailleur}},\
  }\bibfield  {title} {\enquote {\bibinfo {title} {Active brownian particles
  and run-and-tumble particles: A comparative study},}\ }\href@noop {}
  {\bibfield  {journal} {\bibinfo  {journal} {The European Physical Journal
  Special Topics}\ }\textbf {\bibinfo {volume} {224}},\ \bibinfo {pages}
  {1231--1262} (\bibinfo {year} {2015})}\BibitemShut {NoStop}%
\bibitem [{\citenamefont {Fily}\ and\ \citenamefont
  {Marchetti}(2012)}]{fily2012athermal}%
  \BibitemOpen
  \bibfield  {author} {\bibinfo {author} {\bibfnamefont {Y.}~\bibnamefont
  {Fily}}\ and\ \bibinfo {author} {\bibfnamefont {M.~C.}\ \bibnamefont
  {Marchetti}},\ }\bibfield  {title} {\enquote {\bibinfo {title} {Athermal
  phase separation of self-propelled particles with no alignment},}\
  }\href@noop {} {\bibfield  {journal} {\bibinfo  {journal} {Physical review
  letters}\ }\textbf {\bibinfo {volume} {108}},\ \bibinfo {pages} {235702}
  (\bibinfo {year} {2012})}\BibitemShut {NoStop}%
\bibitem [{\citenamefont {Redner}, \citenamefont {Hagan},\ and\ \citenamefont
  {Baskaran}(2013)}]{redner2013structure}%
  \BibitemOpen
  \bibfield  {author} {\bibinfo {author} {\bibfnamefont {G.~S.}\ \bibnamefont
  {Redner}}, \bibinfo {author} {\bibfnamefont {M.~F.}\ \bibnamefont {Hagan}}, \
  and\ \bibinfo {author} {\bibfnamefont {A.}~\bibnamefont {Baskaran}},\
  }\bibfield  {title} {\enquote {\bibinfo {title} {Structure and dynamics of a
  phase-separating active colloidal fluid},}\ }\href@noop {} {\bibfield
  {journal} {\bibinfo  {journal} {Physical review letters}\ }\textbf {\bibinfo
  {volume} {110}},\ \bibinfo {pages} {055701} (\bibinfo {year}
  {2013})}\BibitemShut {NoStop}%
\bibitem [{\citenamefont {Levis}\ and\ \citenamefont
  {Berthier}(2014)}]{levis2014clustering}%
  \BibitemOpen
  \bibfield  {author} {\bibinfo {author} {\bibfnamefont {D.}~\bibnamefont
  {Levis}}\ and\ \bibinfo {author} {\bibfnamefont {L.}~\bibnamefont
  {Berthier}},\ }\bibfield  {title} {\enquote {\bibinfo {title} {Clustering and
  heterogeneous dynamics in a kinetic monte carlo model of self-propelled hard
  disks},}\ }\href@noop {} {\bibfield  {journal} {\bibinfo  {journal} {Physical
  Review E}\ }\textbf {\bibinfo {volume} {89}},\ \bibinfo {pages} {062301}
  (\bibinfo {year} {2014})}\BibitemShut {NoStop}%
\bibitem [{\citenamefont {Fily}, \citenamefont {Henkes},\ and\ \citenamefont
  {Marchetti}(2014)}]{fily2014freezing}%
  \BibitemOpen
  \bibfield  {author} {\bibinfo {author} {\bibfnamefont {Y.}~\bibnamefont
  {Fily}}, \bibinfo {author} {\bibfnamefont {S.}~\bibnamefont {Henkes}}, \ and\
  \bibinfo {author} {\bibfnamefont {M.~C.}\ \bibnamefont {Marchetti}},\
  }\bibfield  {title} {\enquote {\bibinfo {title} {Freezing and phase
  separation of self-propelled disks},}\ }\href@noop {} {\bibfield  {journal}
  {\bibinfo  {journal} {Soft matter}\ }\textbf {\bibinfo {volume} {10}},\
  \bibinfo {pages} {2132--2140} (\bibinfo {year} {2014})}\BibitemShut {NoStop}%
\bibitem [{\citenamefont {Digregorio}\ \emph {et~al.}(2018)\citenamefont
  {Digregorio}, \citenamefont {Levis}, \citenamefont {Suma}, \citenamefont
  {Cugliandolo}, \citenamefont {Gonnella},\ and\ \citenamefont
  {Pagonabarraga}}]{digregorio2018full}%
  \BibitemOpen
  \bibfield  {author} {\bibinfo {author} {\bibfnamefont {P.}~\bibnamefont
  {Digregorio}}, \bibinfo {author} {\bibfnamefont {D.}~\bibnamefont {Levis}},
  \bibinfo {author} {\bibfnamefont {A.}~\bibnamefont {Suma}}, \bibinfo {author}
  {\bibfnamefont {L.~F.}\ \bibnamefont {Cugliandolo}}, \bibinfo {author}
  {\bibfnamefont {G.}~\bibnamefont {Gonnella}}, \ and\ \bibinfo {author}
  {\bibfnamefont {I.}~\bibnamefont {Pagonabarraga}},\ }\bibfield  {title}
  {\enquote {\bibinfo {title} {Full phase diagram of active brownian disks:
  From melting to motility-induced phase separation},}\ }\href@noop {}
  {\bibfield  {journal} {\bibinfo  {journal} {Physical review letters}\
  }\textbf {\bibinfo {volume} {121}},\ \bibinfo {pages} {098003} (\bibinfo
  {year} {2018})}\BibitemShut {NoStop}%
\bibitem [{\citenamefont {Tailleur}\ and\ \citenamefont
  {Cates}(2008)}]{tailleur2008statistical}%
  \BibitemOpen
  \bibfield  {author} {\bibinfo {author} {\bibfnamefont {J.}~\bibnamefont
  {Tailleur}}\ and\ \bibinfo {author} {\bibfnamefont {M.}~\bibnamefont
  {Cates}},\ }\bibfield  {title} {\enquote {\bibinfo {title} {Statistical
  mechanics of interacting run-and-tumble bacteria},}\ }\href@noop {}
  {\bibfield  {journal} {\bibinfo  {journal} {Physical review letters}\
  }\textbf {\bibinfo {volume} {100}},\ \bibinfo {pages} {218103} (\bibinfo
  {year} {2008})}\BibitemShut {NoStop}%
\bibitem [{\citenamefont {Cates}\ and\ \citenamefont
  {Tailleur}(2015)}]{cates2015motility}%
  \BibitemOpen
  \bibfield  {author} {\bibinfo {author} {\bibfnamefont {M.~E.}\ \bibnamefont
  {Cates}}\ and\ \bibinfo {author} {\bibfnamefont {J.}~\bibnamefont
  {Tailleur}},\ }\bibfield  {title} {\enquote {\bibinfo {title}
  {Motility-induced phase separation},}\ }\href@noop {} {\bibfield  {journal}
  {\bibinfo  {journal} {Annu. Rev. Condens. Matter Phys.}\ }\textbf {\bibinfo
  {volume} {6}},\ \bibinfo {pages} {219--244} (\bibinfo {year}
  {2015})}\BibitemShut {NoStop}%
\bibitem [{\citenamefont {Soto}, \citenamefont {Pinto},\ and\ \citenamefont
  {Brito}(2024)}]{PRL}%
  \BibitemOpen
  \bibfield  {author} {\bibinfo {author} {\bibfnamefont {R.}~\bibnamefont
  {Soto}}, \bibinfo {author} {\bibfnamefont {M.}~\bibnamefont {Pinto}}, \ and\
  \bibinfo {author} {\bibfnamefont {R.}~\bibnamefont {Brito}},\ }\bibfield
  {title} {\enquote {\bibinfo {title} {Kinetic theory of motility induced phase
  separation for active brownian particles},}\ }\href {\doibase
  10.1103/PhysRevLett.132.208301} {\bibfield  {journal} {\bibinfo  {journal}
  {Phys. Rev. Lett.}\ }\textbf {\bibinfo {volume} {132}},\ \bibinfo {pages}
  {208301} (\bibinfo {year} {2024})}\BibitemShut {NoStop}%
\bibitem [{\citenamefont {Wang}(2019)}]{wang2019anomalous}%
  \BibitemOpen
  \bibfield  {author} {\bibinfo {author} {\bibfnamefont {J.}~\bibnamefont
  {Wang}},\ }\bibfield  {title} {\enquote {\bibinfo {title} {Anomalous
  diffusion of active brownian particles in crystalline phases},}\ }in\
  \href@noop {} {\emph {\bibinfo {booktitle} {IOP Conference Series: Earth and
  Environmental Science}}},\ Vol.\ \bibinfo {volume} {237}\ (\bibinfo
  {organization} {IOP Publishing},\ \bibinfo {year} {2019})\ p.\ \bibinfo
  {pages} {052005}\BibitemShut {NoStop}%
\bibitem [{\citenamefont {Reichert}\ and\ \citenamefont
  {Voigtmann}(2021)}]{reichert2021tracer}%
  \BibitemOpen
  \bibfield  {author} {\bibinfo {author} {\bibfnamefont {J.}~\bibnamefont
  {Reichert}}\ and\ \bibinfo {author} {\bibfnamefont {T.}~\bibnamefont
  {Voigtmann}},\ }\bibfield  {title} {\enquote {\bibinfo {title} {Tracer
  dynamics in crowded active-particle suspensions},}\ }\href@noop {} {\bibfield
   {journal} {\bibinfo  {journal} {Soft Matter}\ }\textbf {\bibinfo {volume}
  {17}},\ \bibinfo {pages} {10492--10504} (\bibinfo {year} {2021})}\BibitemShut
  {NoStop}%
\bibitem [{\citenamefont {Chapman}\ and\ \citenamefont
  {Cowling}(1990)}]{chapman1990mathematical}%
  \BibitemOpen
  \bibfield  {author} {\bibinfo {author} {\bibfnamefont {S.}~\bibnamefont
  {Chapman}}\ and\ \bibinfo {author} {\bibfnamefont {T.~G.}\ \bibnamefont
  {Cowling}},\ }\href@noop {} {\emph {\bibinfo {title} {The mathematical theory
  of non-uniform gases: an account of the kinetic theory of viscosity, thermal
  conduction and diffusion in gases}}}\ (\bibinfo  {publisher} {Cambridge
  university press},\ \bibinfo {year} {1990})\BibitemShut {NoStop}%
\bibitem [{\citenamefont {Soto}(2016)}]{soto2016kinetic}%
  \BibitemOpen
  \bibfield  {author} {\bibinfo {author} {\bibfnamefont {R.}~\bibnamefont
  {Soto}},\ }\href@noop {} {\emph {\bibinfo {title} {Kinetic theory and
  transport phenomena}}}\ (\bibinfo  {publisher} {Oxford University Press},\
  \bibinfo {year} {2016})\BibitemShut {NoStop}%
\bibitem [{\citenamefont {Van~Beijeren}\ and\ \citenamefont
  {Ernst}(1973)}]{van1973modified}%
  \BibitemOpen
  \bibfield  {author} {\bibinfo {author} {\bibfnamefont {H.}~\bibnamefont
  {Van~Beijeren}}\ and\ \bibinfo {author} {\bibfnamefont {M.~H.}\ \bibnamefont
  {Ernst}},\ }\bibfield  {title} {\enquote {\bibinfo {title} {The modified
  enskog equation},}\ }\href@noop {} {\bibfield  {journal} {\bibinfo  {journal}
  {Physica}\ }\textbf {\bibinfo {volume} {68}},\ \bibinfo {pages} {437--456}
  (\bibinfo {year} {1973})}\BibitemShut {NoStop}%
\bibitem [{\citenamefont {Brilliantov}\ and\ \citenamefont
  {P{\"o}schel}(2004)}]{brilliantov2004kinetic}%
  \BibitemOpen
  \bibfield  {author} {\bibinfo {author} {\bibfnamefont {N.}~\bibnamefont
  {Brilliantov}}\ and\ \bibinfo {author} {\bibfnamefont {T.}~\bibnamefont
  {P{\"o}schel}},\ }\href@noop {} {\emph {\bibinfo {title} {Kinetic theory of
  granular gases}}}\ (\bibinfo  {publisher} {Oxford University Press},\
  \bibinfo {year} {2004})\BibitemShut {NoStop}%
\bibitem [{\citenamefont {Garz{\'o}}(2019)}]{garzo2019granular}%
  \BibitemOpen
  \bibfield  {author} {\bibinfo {author} {\bibfnamefont {V.}~\bibnamefont
  {Garz{\'o}}},\ }\href@noop {} {\emph {\bibinfo {title} {Granular gaseous
  flows}}}\ (\bibinfo  {publisher} {Springer},\ \bibinfo {year}
  {2019})\BibitemShut {NoStop}%
\bibitem [{\citenamefont {Soto}\ and\ \citenamefont
  {Mareschal}(2001)}]{soto2001statistical}%
  \BibitemOpen
  \bibfield  {author} {\bibinfo {author} {\bibfnamefont {R.}~\bibnamefont
  {Soto}}\ and\ \bibinfo {author} {\bibfnamefont {M.}~\bibnamefont
  {Mareschal}},\ }\bibfield  {title} {\enquote {\bibinfo {title} {Statistical
  mechanics of fluidized granular media: Short-range velocity correlations},}\
  }\href@noop {} {\bibfield  {journal} {\bibinfo  {journal} {Physical Review
  E}\ }\textbf {\bibinfo {volume} {63}},\ \bibinfo {pages} {041303} (\bibinfo
  {year} {2001})}\BibitemShut {NoStop}%
\bibitem [{\citenamefont {de~Macedo~Biniossek}\ \emph
  {et~al.}(2018)\citenamefont {de~Macedo~Biniossek}, \citenamefont {L{\"o}wen},
  \citenamefont {Voigtmann},\ and\ \citenamefont {Smallenburg}}]{de2018static}%
  \BibitemOpen
  \bibfield  {author} {\bibinfo {author} {\bibfnamefont {N.}~\bibnamefont
  {de~Macedo~Biniossek}}, \bibinfo {author} {\bibfnamefont {H.}~\bibnamefont
  {L{\"o}wen}}, \bibinfo {author} {\bibfnamefont {T.}~\bibnamefont
  {Voigtmann}}, \ and\ \bibinfo {author} {\bibfnamefont {F.}~\bibnamefont
  {Smallenburg}},\ }\bibfield  {title} {\enquote {\bibinfo {title} {Static
  structure of active brownian hard disks},}\ }\href@noop {} {\bibfield
  {journal} {\bibinfo  {journal} {Journal of Physics: Condensed Matter}\
  }\textbf {\bibinfo {volume} {30}},\ \bibinfo {pages} {074001} (\bibinfo
  {year} {2018})}\BibitemShut {NoStop}%
\bibitem [{\citenamefont {Jeggle}, \citenamefont {Stenhammar},\ and\
  \citenamefont {Wittkowski}(2020)}]{jeggle2020pair}%
  \BibitemOpen
  \bibfield  {author} {\bibinfo {author} {\bibfnamefont {J.}~\bibnamefont
  {Jeggle}}, \bibinfo {author} {\bibfnamefont {J.}~\bibnamefont {Stenhammar}},
  \ and\ \bibinfo {author} {\bibfnamefont {R.}~\bibnamefont {Wittkowski}},\
  }\bibfield  {title} {\enquote {\bibinfo {title} {Pair-distribution function
  of active brownian spheres in two spatial dimensions: simulation results and
  analytic representation},}\ }\href@noop {} {\bibfield  {journal} {\bibinfo
  {journal} {The Journal of Chemical Physics}\ }\textbf {\bibinfo {volume}
  {152}} (\bibinfo {year} {2020})}\BibitemShut {NoStop}%
\bibitem [{\citenamefont {Henderson}(1975)}]{henderson1975simple}%
  \BibitemOpen
  \bibfield  {author} {\bibinfo {author} {\bibfnamefont {D.}~\bibnamefont
  {Henderson}},\ }\bibfield  {title} {\enquote {\bibinfo {title} {A simple
  equation of state for hard discs},}\ }\href@noop {} {\bibfield  {journal}
  {\bibinfo  {journal} {Molecular Physics}\ }\textbf {\bibinfo {volume} {30}},\
  \bibinfo {pages} {971--972} (\bibinfo {year} {1975})}\BibitemShut {NoStop}%
\bibitem [{\citenamefont {Boon}\ and\ \citenamefont
  {Yip}(1991)}]{boon1991molecular}%
  \BibitemOpen
  \bibfield  {author} {\bibinfo {author} {\bibfnamefont {J.~P.}\ \bibnamefont
  {Boon}}\ and\ \bibinfo {author} {\bibfnamefont {S.}~\bibnamefont {Yip}},\
  }\href@noop {} {\emph {\bibinfo {title} {Molecular hydrodynamics}}}\
  (\bibinfo  {publisher} {Courier Corporation},\ \bibinfo {year}
  {1991})\BibitemShut {NoStop}%
\bibitem [{\citenamefont {Bialk{\'e}}, \citenamefont {L{\"o}wen},\ and\
  \citenamefont {Speck}(2013)}]{bialke2013microscopic}%
  \BibitemOpen
  \bibfield  {author} {\bibinfo {author} {\bibfnamefont {J.}~\bibnamefont
  {Bialk{\'e}}}, \bibinfo {author} {\bibfnamefont {H.}~\bibnamefont
  {L{\"o}wen}}, \ and\ \bibinfo {author} {\bibfnamefont {T.}~\bibnamefont
  {Speck}},\ }\bibfield  {title} {\enquote {\bibinfo {title} {Microscopic
  theory for the phase separation of self-propelled repulsive disks},}\
  }\href@noop {} {\bibfield  {journal} {\bibinfo  {journal} {Europhysics
  Letters}\ }\textbf {\bibinfo {volume} {103}},\ \bibinfo {pages} {30008}
  (\bibinfo {year} {2013})}\BibitemShut {NoStop}%
\bibitem [{\citenamefont {de~Pirey}, \citenamefont {Lozano},\ and\
  \citenamefont {Van~Wijland}(2019)}]{de2019active}%
  \BibitemOpen
  \bibfield  {author} {\bibinfo {author} {\bibfnamefont {T.~A.}\ \bibnamefont
  {de~Pirey}}, \bibinfo {author} {\bibfnamefont {G.}~\bibnamefont {Lozano}}, \
  and\ \bibinfo {author} {\bibfnamefont {F.}~\bibnamefont {Van~Wijland}},\
  }\bibfield  {title} {\enquote {\bibinfo {title} {Active hard spheres in
  infinitely many dimensions},}\ }\href@noop {} {\bibfield  {journal} {\bibinfo
   {journal} {Physical review letters}\ }\textbf {\bibinfo {volume} {123}},\
  \bibinfo {pages} {260602} (\bibinfo {year} {2019})}\BibitemShut {NoStop}%
\bibitem [{\citenamefont {Akintunde}\ \emph {et~al.}(2024)\citenamefont
  {Akintunde}, \citenamefont {Bayati}, \citenamefont {Row},\ and\ \citenamefont
  {Mallory}}]{akintunde2024single}%
  \BibitemOpen
  \bibfield  {author} {\bibinfo {author} {\bibfnamefont {A.}~\bibnamefont
  {Akintunde}}, \bibinfo {author} {\bibfnamefont {P.}~\bibnamefont {Bayati}},
  \bibinfo {author} {\bibfnamefont {H.}~\bibnamefont {Row}}, \ and\ \bibinfo
  {author} {\bibfnamefont {S.~A.}\ \bibnamefont {Mallory}},\ }\bibfield
  {title} {\enquote {\bibinfo {title} {Single-file diffusion of active brownian
  particles},}\ }\href@noop {} {\bibfield  {journal} {\bibinfo  {journal}
  {arXiv preprint arXiv:2411.08988}\ } (\bibinfo {year} {2024})}\BibitemShut
  {NoStop}%
\bibitem [{\citenamefont {Dulaney}, \citenamefont {Mallory},\ and\
  \citenamefont {Brady}(2021)}]{dulaney2021isothermal}%
  \BibitemOpen
  \bibfield  {author} {\bibinfo {author} {\bibfnamefont {A.~R.}\ \bibnamefont
  {Dulaney}}, \bibinfo {author} {\bibfnamefont {S.~A.}\ \bibnamefont
  {Mallory}}, \ and\ \bibinfo {author} {\bibfnamefont {J.~F.}\ \bibnamefont
  {Brady}},\ }\bibfield  {title} {\enquote {\bibinfo {title} {The
  “isothermal” compressibility of active matter},}\ }\href@noop {}
  {\bibfield  {journal} {\bibinfo  {journal} {The Journal of Chemical Physics}\
  }\textbf {\bibinfo {volume} {154}} (\bibinfo {year} {2021})}\BibitemShut
  {NoStop}%
\end{thebibliography}
\end{document}